\documentclass[]{emulateapj}
\begin{document}
\title{A Million-Second Chandra View of Cassiopeia A}


\author{Una Hwang\altaffilmark{1,2}, J. Martin Laming\altaffilmark{3},
Carles Badenes\altaffilmark{4}, Fred Berendse\altaffilmark{5}, John
Blondin\altaffilmark{6}, Denis Cioffi\altaffilmark{7}, Tracey
DeLaney\altaffilmark{8}, Daniel Dewey\altaffilmark{9}, Robert
Fesen\altaffilmark{10}, Kathryn A.  Flanagan\altaffilmark{9},
Christopher L. Fryer\altaffilmark{11}, Parviz
Ghavamian\altaffilmark{12}, John P. Hughes\altaffilmark{13}, Jon
A. Morse\altaffilmark{14}, Paul P. Plucinsky\altaffilmark{15}, Robert
Petre\altaffilmark{1}, Martin Pohl\altaffilmark{16}, Lawrence
Rudnick\altaffilmark{8}, Ravi Sankrit\altaffilmark{12}, Patrick
O. Slane\altaffilmark{15}, Randall K. Smith\altaffilmark{1,12}, Jacco
Vink\altaffilmark{17}, Jessica S. Warren\altaffilmark{13}}
\altaffiltext{1}{Code 662, NASA GSFC; hwang, rob, rsmith@milkyway.gsfc.
nasa.gov}
\altaffiltext{2}{University of Maryland}
\altaffiltext{3}{Code 7674L, Naval Research Laboratory; jlaming@ssd5.nrl.
navy.mil}
\altaffiltext{4}{Institut d'Estudis Espacials de Catalunya, Espa\~{n}a; badenes@
ieec.fcr.es}
\altaffiltext{5}{Code 7655.7, Naval Research Laboratory; fberendse@ssd5.
nrl.navy.mil}
\altaffiltext{6}{North Carolina State University; John\_Blondin@ncsu.edu}
\altaffiltext{7}{The George Washington University; professor@cioffi.us}
\altaffiltext{8}{University of Minnesota; tdelaney, larry@astro.umn.edu}
\altaffiltext{9}{Center for Space Research, MIT; dd, kaf@space.mit.edu}
\altaffiltext{10}{Dartmouth College; fesen@snr.dartmouth.edu}
\altaffiltext{11}{Los Alamos National Laboratory; fryer@lanl.gov}
\altaffiltext{12}{The Johns Hopkins University; parviz, ravi@pha.jhu.edu}
\altaffiltext{13}{Rutgers University; jackph, jesawyer@physics.rutgers.edu}
\altaffiltext{14}{Arizona State University; jon.morse@asu.edu}
\altaffiltext{15}{Harvard-Smithsonian Center for Astrophysics; plucinsk, slane@head.cfa.harvard.edu}
\altaffiltext{16}{Iowa State University; mkp@iastate.edu}
\altaffiltext{17}{SRON National Institute for Space Research; j.vink@sron.nl}

\begin{abstract}
We introduce a million-second observation of the supernova remnant
Cassiopeia A with the Chandra X-ray Observatory.  The bipolar
structure of the Si-rich ejecta (NE jet and SW counterpart) is clearly
evident in the new images, and their chemical similarity is confirmed
by their spectra.  These are most likely due to jets of ejecta as
opposed to cavities in the circumstellar medium, since we can reject
simple models for the latter. The properties of these jets and the
Fe-rich ejecta will provide clues to the explosion of Cas~A.
\end{abstract}

\section{Introduction}

The young core-collapse supernova remnant Cassiopeia A is a unique
astrophysical laboratory for studying ejecta and shocks in remnants
because of its brightness across the electromagnetic spectrum and its
well-constrained age and distance. Earlier X-ray observations with
the Chandra Observatory have already produced significant results,
such as the discovery of the compact stellar remnant (Tananbaum et
al. 1999), identification of major types of nucleosynthesis products
(Hughes et al. 2000, Laming \& Hwang 2003, Hwang \& Laming 2003),
measurement of magnetic fields (Vink \& Laming 2003), identification
of the forward and reverse shocks (Gotthelf et al. 2001), and
measurement of small-scale proper motions (DeLaney et al. 2003,
2004). The remnant is believed to have lost most of its progenitor
mass prior to explosion at $\sim$ 4 M$_\odot$ (e.g., Willingale et
al. 2003, Chevalier \& Oishi 2003, Laming \& Hwang 2003).

A 1 Ms observation of Cas~A was performed as a Very Large Project
(VLP) in Chandra Cycle 5. It was largely motivated by two papers
(Laming \& Hwang 2003; Hwang \& Laming 2003; hereafter LH03 and HL03),
which used Chandra's subarcsecond angular resolution to isolate
compact ejecta features in Cas~A.  Since the reverse shock crosses
these knots in a very short time compared to the age of the remnant,
the extracted knot spectra can be characterized with a single electron
temperature and ionization age. These are then straightforwardly
compared to model results to infer details of the ejecta density
profile, the explosion asymmetry, and with further modest assumptions,
the Lagrangian mass coordinates of ejecta structures.  With earlier
Chandra observations, individual ejecta knots could not always be
isolated when spectra were extracted over several image
resolution elements. A 1 Ms exposure allows nearly every pixel of the
ACIS CCD image of Cas~A to provide a spectrum for
fitting. Revisiting the work of LH03 and HL03 with a larger selection
of Cas~A ejecta knots would be the first full analysis of its kind and
a significant effort that is well beyond the scope of this letter.  In
this introductory paper, we present new images, highlight some results
immediately revealed by the deep pointing, and indicate areas of
future work where significant findings are expected.

\section{Observations and Data Reduction}

Cas~A was observed for 1 Ms with the backside illuminated S3 chip of
the Advanced CCD Imaging Spectrometer (ACIS) on the Chandra
Observatory.  The first 50 ks segment was completed between 8 and 9
February 2004, with the same pointing, frame time (3.2 s), and roll
angle (325$^\circ$) as for two earlier 50 ks ACIS observations to
facilitate proper motion studies.  The remaining 950 ks were executed
in eight segments between 14 April and 6 May 2004 with a 3.0 s frame
time and no roll angle constraints (the roll angles drifted from
40$^\circ$ to 65$^\circ$).  The aimpoint in the first 50 ks segment
(and in the earlier 50 ks observations) is roughly 1.9$'$ SE of the
point source, while it is 2.4$'$ NE in the remaining segments.
Cas~A's high count rate requires the use of only a single CCD chip and
accumulation of the data in GRADED mode to avoid telemetry loss.  The
spectra thus cannot be corrected to reduce the effects of charge
transfer inefficiency.
%, which increases with the age of the mission.
%(We find roughly 10-20\% differences in spectral parameters derived
%for a sample of knots in Cas~A using Cycle 5 data as compared to Cycle
%1 data.)

We retain only events with energies from 0.3-10 keV and apply
time-dependent gain corrections for the S3 chip.  Time segments with
background flares were excluded by making consistent count-rate cuts
based on the off-source light curve.  The filtered data set contained
282 million photons over the 5$'$ extent of Cas~A and had an exposure
time of 980 ks.

\section{Results}
%\subsection{Overview}

\noindent{\bf Overview:} A three-color composite of Cas~A in Si He
$\alpha$ (1.78-2.0 keV; red), Fe K (6.52-6.95 keV; blue) and 4.2-6.4
keV continuum emission (green) is shown in Figure 1a.  The images have
not been smoothed nor the continuum contribution removed (i.e., the
images contain both line and continuum photons).  The figure shows
that the filamentary 4-6 keV continuum emission is fragmented, but
circular overall, and present throughout the remnant image (see also
Gotthelf et al. 2001). Most of this emission is probably synchrotron
emission generated at the forward shock.  The reverse-shocked ejecta
emission, by contrast, is highly non-circular, with the prominent
linear jets of Si-rich ejecta (in red) in the northeast (NE), their
fainter counterparts to the southwest (SW), and Fe-rich ejecta fingers
(in blue) extending to the forward shock in the southeast (see also
Hughes et al. 2000, Hwang et al. 2000).
%substitute figure, text from LR?

\begin{figure*}
\includegraphics[scale=1.00]{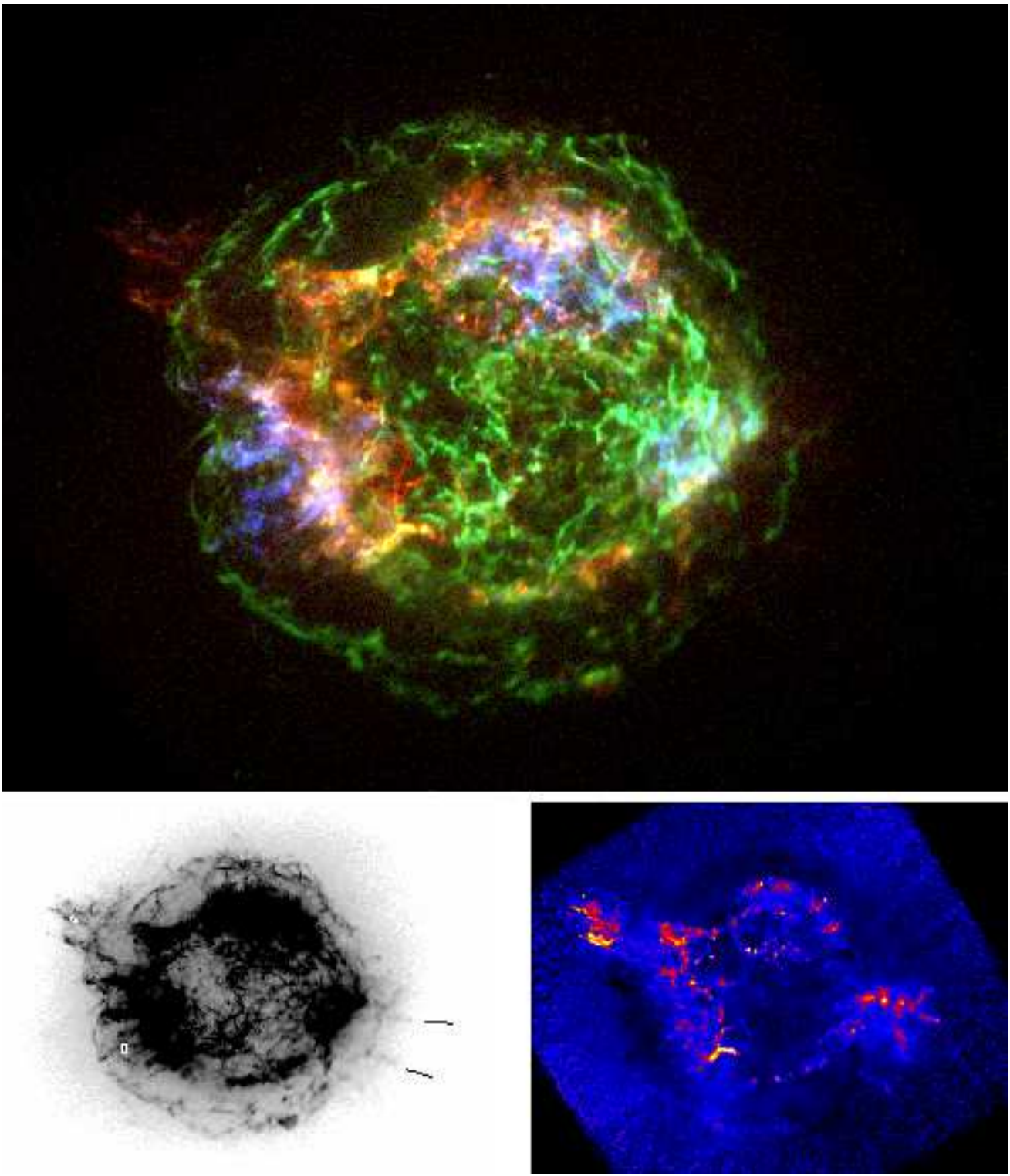}
\caption{(Top) Three-color image of Cas~A with red=Si He $\alpha$
(1.78-2.0 keV), blue=Fe K (6.52-6.95 keV), green=4.2-6.4 keV
continuum.  The remnant is roughly $5'$ across. (Bottom left)
Over-exposed broadband image showing faint features.  The spectral
regions are indicated (upper left box: NE jet, lower left box: Fe-rich
region, lines at lower right point to two SW jet filaments).
%The lower SW jet filament is nearly perfectly aligned with the point source and the longest NE jet filament. 
Smearing associated with CCD readout causes the low-surface brightness
artifacts outside the remnant to the SE and NW. (Bottom right) On the
same scale, the ratio image of the Si He $\alpha$ (1.78-2.0 keV) and
1.3-1.6 keV (Mg He $\alpha$, Fe L), without subtraction of the
continuum contribution.  The image highlights the jet and counterjet
traced by Si emission, though features at the lowest intensity levels
are uncertain.}
%The point source is visible in this image, and is clearly {\it offset} relative to the jets.}
\end{figure*}

The image revealed several faint point sources around the remnant
which matched the positions of optical stars, thereby permitting
accurate coordinate transformation of the image relative to the USNO
A2.0 catalog.  The resulting measured position of the remnant's
central X-ray point source was found to be (J2000) $\alpha$=23h 23m
27.945s$\pm$0.08s, $\delta$ = 58$^\circ$ 48$'$ 42.45$'' \pm 0.6''$
(see Fesen, Pavlov, \& Sanwal 2004, in preparation).

%A close-up view of the northeastern jet filaments demonstrates the image
%quality of the deep observation, with Figure 2 comparing the broadband
%Cycle 1 and VLP images. 
The overexposed broadband image shown in Figure 1b highlights faint
features in Cas~A, including the jet structures in the NE and SW.  An
X-ray jet counterpart in the SW had been hinted in earlier Chandra
observations (Hwang et al. 2000), but is clearly evident in the new
image.  It is further highlighted in the ratio image of Si XIII He
$\alpha$ (plus continuum, 1.78-2.0 keV) relative to the 1.3-1.6 keV
band (mostly weak Mg XI and Fe L plus continuum) shown in Figure 1c
(see also Vink 2004). The SW jet filaments are fainter, have a larger
opening angle on the sky, and do not extend as far inward as their NE
counterparts.  Though the point source is clearly offset relative to
the jets as a whole (Figure 1c), it is well-aligned with the longest
NE filament and the lower of two straight SW filaments marked in
Figure 1b.  A bipolar ejecta structure is already known from optical
spectral observations (Fesen 2001) that reveal high-velocity
($>$10,000 km/s) ejecta knots beyond the forward shock in the jet
regions, and is now also seen in 24$\mu$m dust emission (Hines et
al. 2004).
%Exact correspondence between the optical or infrared knots and the X-ray
%knots is not expected.

The chemical similarity of the jets is demonstrated by their spectra,
shown in Figure 2a. Both the spectrum from a filament in the NE jet and
the summed spectrum of two SW jet filaments (all at similar radial
distance 3.3$'$ from the compact source) show strong Si, S, Ar, and Ca
emission, as well as a Fe K blend.  Their thermal characteristics are
also similar, with preliminary spectral fits to simple models with an
underlying O continuum giving temperatures kT of $\sim 1-1.5$ keV and
ionization ages $n_et$ a few times $10^{11}$ cm$^{-3}$s.  These knots
resemble the typical Si/O-rich knots studied by LH03.

\begin{figure}
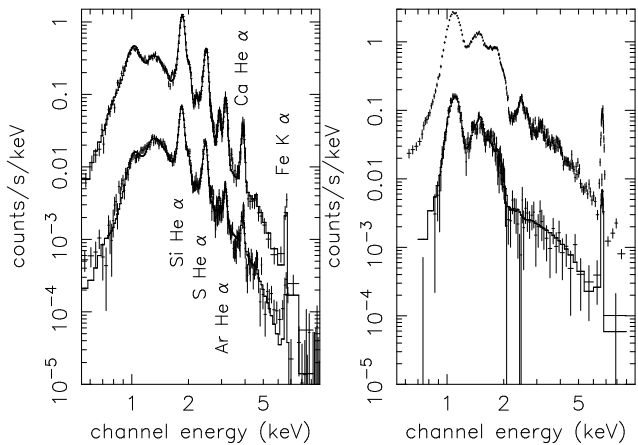

\epsscale{0.5}
\includegraphics[scale=0.36,angle=-90]{fig2a.ps}\includegraphics[scale=0.36,angle=-90]{fig2b.ps}
\caption{(left) Comparison of spectrum of a filament in the NE jet
scaled by a factor of 15 and the summed spectrum of two filaments in
the SW counterjet.  The filaments are at roughly the same
radius, and both display strong Si, S, Ca, and Ar lines, and a modest
Fe K line.  (right) Comparison of Fe cloud spectrum in summed 50 ks
ACIS observations of January 2000 and 2002 (scaled by 15) and VLP ACIS
observations from 2004, confirming the relative purity of the Fe, seen
in both line and continuum emission.}
\end{figure}

In the southeast, where extremely Fe-rich material has been observed,
the spectrum of the highly Fe-rich cloud discussed in HL03 is
confirmed by the VLP spectrum shown in Figure 2b to be virtually pure
Fe. The spectrum is dominated by Fe L and Fe K line blends and Fe
continuum.  Highly Fe-rich ejecta are sites of complete Si burning in
the explosion, and may also harbor some of the $^{44}$Ti known to have
been synthesized in $\alpha$-rich freeze out (Iyudin et al. 1994, Vink
et al. 2001, Vink \& Laming 2003). As such, they originated close to
the explosion center and are very sensitive to the details of the
explosion. HL03 estimate that a few percent of the total Fe ejected by
the explosion is currently visible in X-rays, which is similar to the
%while most of the rest is probably still interior to the reverse shock. This fraction is
fraction of Ni inferred to have been mixed out into the envelope of SN
1987A by Rayleigh-Taylor instabilities during its explosion (Pinto \&
Woosley 1988). Identification and modelling of other Fe-rich regions
will be the focus of future work, and will offer the best comparisons
with theoretical models for Fe-production.

%\begin{figure}
%{\includegraphics[scale=0.18,angle=-90]{f3a.ps}\includegraphics[scale=0.18,angle=-90]{f3b.ps}}
%\end{figure}

%\subsection{The Nature of the Jets}
\noindent{\bf The Nature of the Jets:} Jet structures can result from an
explosion into local cavities of the interstellar medium formed by
asymmetric winds in the late evolutionary phases of the progenitor
star (Blondin et al. 1996). However, models of the shock trajectories
by Truelove and McKee (1999) as implemented by LH03 (but for lower CSM
density), show that the forward shock decelerates very slowly in a
local cavity.  The reverse shock is thus not strong enough to progress
through enough ejecta to yield the relatively high ionization age
observed for the plasma in the jet-like region of Cas~A. We
tentatively conclude that Cas A was formed by an asymmetric explosion
that produced jets of ejecta, though more detailed work is required
for confirmation.
This conclusion is corroborated by the chemical enrichment and high
expansion velocities of the optically-emitting jet knots (Fesen 2001).
Since the optical ejecta knots are generally accepted as being
minimally decelerated by the circumstellar medium, an origin in the
explosion mechanism is suggested for the jet knots, which have
significantly higher velocities than elsewhere.

Interesting questions then arise as to the nature and energetics of
the jets. A jet-induced explosion (e.g. Khokhlov et al. 1999) requires
an overdense jet to explode the star and is expected to leave large
amounts of Fe ejecta in the jet regions. This contradicts existing
X-ray and optical observations of Cas~A (e.g. Hughes et al. 2000,
Fesen 2001).  Furthermore, the kick direction for the compact object
that is inferred from the motions of optical knots (Thorstensen et
al. 2001) is not {\it along} the jet axis as would be expected, but is
instead perpendicular to it.  Alternatively, an underdense
relativistic jet could be driven by a collapsar (see Zhang, Woosley \&
Heger 2004, Nagataki et al. 2003). In this scenario, the ejecta knots
observed in the jet regions could be
%should have an origin in the ambient ejecta material that entrains the jet, and
pulled off 
at the interface of the jet and ambient ejecta plasma and are unlikely
to be Fe-rich.  Collapsars are believed to produce super-energetic
explosions (i.e., $\gamma$-ray bursts), but estimates of the explosion
energy for Cas~A by LH03 at $2-4 \times 10^{51}$\,erg suggest that it
was closer to a normal supernova, and therefore unlikely to have
actually produced a $\gamma$-ray burst (Podsiadlowski et al. 2004).

Asymmetries in normal supernova explosions are much milder (Fryer \&
Heger 2000, Scheck et al. 2004).  Whether these explosions can produce
the observed asymmetry depends sensitively on the exploding star.
Asymmetries are effectively erased if the star retains much of its
hydrogen envelope (see Chevalier \& Soker 1989 for the case of 1987A).
Loss of the hydrogen envelope could allow asymmetries to persist
through break-out, and this may explain the jet-like structures in Cas
A, since it was much less massive at explosion than the progenitor of
SN 1987A.  Though these arguments are speculative, we expect to reach
much firmer conclusions through spectroscopy of faint jet knots with
the VLP data and by coupling hydrodynamics with a treatment of the
ionization balance and electron heating.

\noindent{\bf Point Source:} The deep image gives no evidence for an
extended pulsar wind nebula surrounding the point source.  As
ray-tracing simulations confirm, the slightly extended appearance of
the point-source image can be explained by the change in the slightly
off-axis position of the point source caused by the varying spacecraft
roll angle in each individual observation segment.

Detailed analysis of the compact source, including spectra and
timing, will be presented in a future paper. Here we simply note that
the best single-component model is a blackbody ($\chi^2 = 493.7$ for
315 degrees of freedom) for $T_{bb}^\infty = (4.89 \pm 0.07) \times
10^6$~K, an emitting surface with radius $R_{bb}^\infty = 0.83 \pm
0.03$~km and an absorbing column of $N_H = (1.25 \pm 0.03) \times
10^{22}{\rm\ cm}^{-2}$.  As in previous studies, the temperature is
higher than expected from cooling models for a $\sim 330$ yr old
neutron star, and the inferred size of the emitting region is small.
Emission from discrete hot spots on the stellar surface may be
indicated (Pavlov et al. 2000, Chakrabarty et al. 2001).

\section{Conclusions}

A new million-second observation of the supernova remnant Cassiopeia A
has been obtained with the S3 CCD detector on the Chandra X-ray
Observatory.
The ejecta have a bipolar structure, implying jets that were formed by
the explosion. For such asymmetries to survive beyond the outer layers
of the star, the explosion at its source must have an extreme
collimation currently only found in hypernova models.  Relativistic
jets may indeed be indicated by the lack of appreciable Fe in their
X-ray spectra, but the total inferred explosion energy for Cas~A is
closer to that of a normal supernova.  Cas~A thus provides good
evidence that jets
may also be produced by normal supernovae.  

The Fe ejecta observed elsewhere in the remnant hold clues to the
nucleosynthesis that took place during the explosion.  Addressing
these kinds of questions with detailed modelling and analysis of these
data should provide insights into the explosion of Cas~A.

\acknowledgements We thank Alexei Khokhlov for scientific discussion,
and the referee for helpful suggestions.  This work is supported by a
grant from the CXC GO program; JML was also supported by basic
research funds of the Office of Naval Research. For more information
about the Cas A VLP, see
http://lheawww.gsfc.nasa.gov/users/hwang/casavlp.html.

\end{document}